\documentstyle[aps,twocolumn,prl,epsf]{revtex}

\begin{document}

\newcommand{\be}{\begin{equation}}
\newcommand{\ee}{\end{equation}}
\newcommand{\bn}{\begin{eqnarray}}
\newcommand{\en}{\end{eqnarray}}

\draft

\twocolumn[\hsize\textwidth\columnwidth\hsize\csname @twocolumnfalse\endcsname

\title{Orbital Switching and the First-Order Insulator-Metal 
Transition in \\ Paramagnetic $V_{2}O_{3}$}

\author{M. S. Laad, L. Craco and E. M\"uller-Hartmann}

\address{Institut f\"ur Theoretische Physik, Universit\"at zu K\"oln, 
77 Z\"ulpicher Strasse, D-50937 K\"oln, Germany \\}
\date{\today}
\maketitle

\widetext

\begin{abstract}
The first-order metal-insulator transition (MIT) in paramagnetic $V_{2}O_{3}$ 
is studied within the ab-initio scheme LDA+DMFT, which merges the local 
density approximation (LDA) with dynamical mean field theory (DMFT).  With a 
fixed value of the Coulomb $U=6.0~eV$, we show how the abrupt pressure driven 
MIT is understood in a new picture: pressure-induced decrease of the trigonal 
distortion within the strong correlation scenario (which is not obtained 
within LDA). We find good quantitative agreement with $(i)$ switch of the 
orbital occupation of $(a_{1g},e_{g1}^{\pi}, e_{g2}^{\pi})$ and the spin 
state $S=1$ across the MIT, $(ii)$ thermodynamics and $dc$ resistivity, and 
$(iii)$ the one-electron spectral function, within this new scenario.
\end{abstract}


]

\narrowtext

The spectacular metal-insulator transition (MIT) in the paramagnetic (P) state
in $V_{2}O_{3}$ has long been considered as a classic, and by now almost a 
textbook version of correlation-driven MIT in a one-band correlated system
described by the one-band Hubbard model~\cite{[1]}.  This conventional wisdom 
has been recently challenged by new experiments, clearly showing the coupled
spin-orbital character of the system, and necessitating a revision in terms of
a multiband picture.  

The conventionally accepted picture rested upon the following argument. In the 
high-$T$ phase, $V_{2}O_{3}$ exists in the corundum structure, and with a 
3$d^{2}$ state ($V^{3+}$), the two $e_{g}^{\sigma}$ orbitals are empty, while
the triply degenerate $t_{2g}$ orbitals are filled by two electrons.  This 
triple degeneracy is lifted by a small trigonal distortion, leading to a singly
occupied $a_{1g}$ orbital oriented along the $c$-axis, and doubly degenerate
planar $e_{g}^{\pi}$ orbitals, occupied by the second electron.  Castellani 
{\it et al.}~\cite{[2]} proposed that strong covalent effects lead to a 
bonding singlet $a_{1g}$ state involving two $V$ ions along the $c$-axis. The 
remaining electron in the two-fold degenerate $e_{g}^{\pi}$ orbitals gives 
rise to a $S=1/2$ model with orbital degeneracy.  This inspired development 
of theoretical techniques, culminating in the dynamical mean field theory 
(DMFT)~\cite{[3]}, leading to considerable improvement in our understanding 
of the MIT.  Within the basic picture~\cite{[2]}, Rozenberg 
{\it et al.}~\cite{[4]} used DMFT for one- and two-orbital Hubbard models 
with Bethe density-of-states to study the MIT in $V_{2}O_{3}$.

Recent polarized X-ray scattering results of Park {\it et al.}~\cite{[5]}
require, however, an interpretation in terms of a spin $S=1$ at each $V$ 
site, with a mixed orbital $e_{g}^{\pi}a_{1g}:e_{g}^{\pi}e_{g}^{\pi}=x:(1-x)$ 
configuration. An exciting conclusion from these results is that the above 
ratio changes its value {\it abruptly} at the MIT, forcing one to abandon the 
one-band Hubbard model to describe $V_{2}O_{3}$.  Notice that this implies an 
important role for the trigonal splitting, since the lower-lying orbital will 
be more ``localized'' when local Coulomb interactions are switched on. This 
raises questions concerning a possible link between the orbital ``switching'' 
and the drastic change in the electronic state, and to a possible common 
underlying origin.

The high-spin ground state results from strong Hund coupling, a fact borne 
out by LDA+U~\cite{[6]}.  Furthermore, LDA+U results argue in favor of models 
without orbital degeneracy.  This line of thought has been extended in detail
by Tanaka~\cite{[7]}, including spin-orbit coupling effects in a cluster-based
approach. Mila {\it et al.}~\cite{[8]} have proposed a model with $S=1$ 
{\it and} orbital degeneracy for the antiferromagnetic insulating phase. 

Hallmarks of strong electronic correlations in $V_{2}O_{3}$ have been observed
across the MIT in thermodynamic~\cite{[9]}, $dc$ transport~\cite{[10]}, 
photoemission~\cite{[11]}, and optical~\cite{[12]} responses.  In particular,
specific heat measurements give $\gamma_{el}\simeq 4.4$, 
$\rho_{dc}(T)=\rho_{0}+AT^{2}$ with large $A$ in the PM phase, while 
PES and optical measurements clearly reveal gap formation in the PI phase,
{\it and} the development of low-energy quasicoherent weight in the PM, with 
a dynamical spectral weight transfer (SWT) from high- to low energy over a 
wide energy scale ($\simeq 4~eV$).  This type of dynamical SWT-driven MIT is 
beyond the scope of LDA+U, which treats dynamical effects of local Coulomb 
interactions within static Hartree-Fock (HF) approximation. These are 
precisely the effects captured reliably by DMFT. 

Thus, a consistent description of the MIT, along with an understanding of
the strong correlation features requires a combination of structural aspects
of $V_{2}O_{3}$ (encoded in LDA+U) with a reliable many-body theory like DMFT.
In this letter, we study these questions within LDA+DMFT~\cite{[13]}, which has
been shown to provide a good quantitative {\it ab-initio} description of
correlated electronic systems.  Moreover the MIT in the P-phase has been 
studied~\cite{[13]} using LDA+DMFT(QMC).  However, the link between the MIT 
and the abrupt switching of orbital occupation has, to our knowledge, not been
explored in detail.  We should mention that such a scenario may have broader 
application to other systems, notably in $Ca_{2-x}Sr_{x}RuO_{4}$~\cite{[14]},
$\;$ and $\;$ is $\;$  a $\;$ hallmark $\;$ of $\;$ the  
\begin{figure}[htb]
\epsfxsize=3.5in
\epsffile{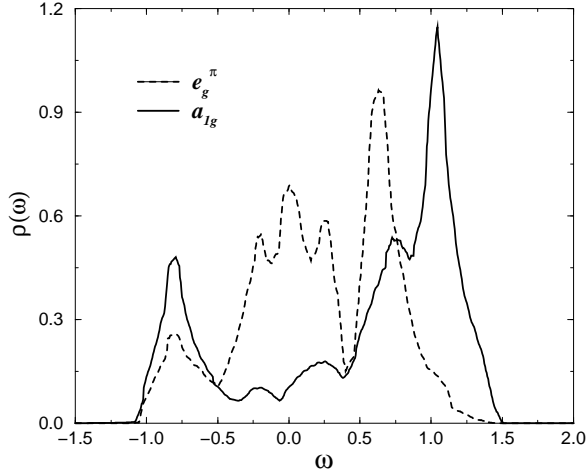}
\caption{ LDA partial density of states for the $e_{g}^{\pi}$ (dashed) 
and $a_{1g}$ (solid) orbitals, obtained from Ref.~[13].} 
\label{fig1}
\end{figure}
\hspace{-0.35cm}importance of orbital correlations in a system.  

We start with the actual LDA bandstructure of $V_{2}O_{3}$ in the corundum 
structure~\cite{[13]}. These results represent (Fig.~\ref{fig1}) metallic 
behavior, with comparable width $(2.5~eV)$ for both $a_{1g}$ and 
$e_{g}^{\pi}$ bands, and a strongly asymmetric structure. With an onsite 
$U\simeq 5-6~eV$, this would invalidate a MO-based approach~\cite{[6]}.  
Further from the $a_{1g}-e_{g}^{\pi}$ splitting, the trigonal field splitting 
is estimated to be $\simeq 0.4~eV$.  The LDA densities of states with and 
without the trigonal distortion do not differ much, and in fact are both 
representative of good metallic behavior, showing that the MIT cannot be 
related purely to $\Delta_{trg}$.  Notice that at the local level, external 
pressure decreases $\Delta_{trg}$, resulting in a sudden increase in 
$n_{a_{1g}}$  around $\Delta_{trg}=0.28~eV$~\cite{[7]}. Turning on the local 
interactions, $U, J_{H}$ and $U'=(U-2J_{H})$ will push up most of the spectral 
weight to the Hubbard sub-bands, and strong $J_{H}$ will favor $S=1$ at each 
$V$ site.  So, while these ``ground state'' features are reasonably well 
accounted for by LDA(+U), aspects like the mixed orbital configuration, the 
switch in orbital occupation at the MIT accompanied by dynamical SWT require 
a reliable treatment of dynamics of correlated electrons. 

Thus, the one-particle part (LDA) of the Hamiltonian is
\be
H_{0}=\sum_{{\bf k}\alpha\beta\sigma}\epsilon_{\alpha\beta}({\bf k})
c_{{\bf k}\alpha\sigma}^{\dag}c_{{\bf k}\beta\sigma}
\ee
with the total DOS, $\rho_{total}^{0}(\epsilon)=\sum_{{\bf k}\alpha\beta\sigma}
\delta(\epsilon-\epsilon_{\alpha\beta}({\bf k}))$, and 
$\alpha,\beta=a_{1g},e_{g1}^{\pi}, e_{g2}^{\pi}$.  To avoid double counting 
of interactions which are already treated on the average by LDA, we 
follow~\cite{[13]} to write

\be
H_{LDA}^{0}=\sum_{{\bf k}\alpha\beta\sigma}\epsilon_{\alpha\beta}({\bf k})
c_{{\bf k}\alpha\sigma}^{\dag}c_{{\bf k}\beta\sigma} + \sum_{i\alpha\sigma}
\epsilon_{i\alpha\sigma}^{0}n_{i\alpha\sigma},
\ee
where $\epsilon_{i\alpha\sigma}^{0}=\epsilon_{i\alpha\sigma}-
U(n_{\alpha\sigma}-\frac{1}{2})+\frac{1}{2}J_{H}(n_{\alpha \bar\sigma}-1)$, 
with $U, J_{H}$ as defined below.
    
With the interactions in the $t_{2g}$ sector, the full Hamiltonian reads,

\be
H=H_{0}+U\sum_{i\alpha}n_{i\alpha\uparrow}n_{i\alpha\downarrow} + 
\sum_{i\alpha\beta\sigma\sigma'}U_{\alpha\beta}^{\sigma\sigma'}
n_{i\alpha\sigma}n_{i\beta\sigma'} \;. 
\ee
Constrained LDA calculations yield $U=5-6~eV$ (without inclusion of screening
of $t_{2g}$ interactions by $e_{g}$ electrons), $J_{H}\simeq 1~eV$, and 
$U_{\alpha\beta}^{\sigma\sigma'}\equiv U'=(U-2J_{H})=3-4~eV$.  
Following~\cite{[13]}, we use the fact that the $e_{g}^{\sigma}$ bands are 
well separated from the $t_{2g}$ bands to consider only the $t_{2g}$ manifold.
Furthermore, in the PM and para-orbital phase, we have 
$G_{\alpha\beta\sigma\sigma'}(\omega)=\delta_{\alpha\beta}
\delta_{\sigma\sigma'}G_{\alpha\sigma}(\omega)$ and 
$\Sigma_{\alpha\beta\sigma\sigma'}(\omega)=\delta_
{\alpha\beta}\delta_{\sigma\sigma'}\Sigma_{\alpha\sigma}(\omega)$.

In the $t_{2g}$ sub-basis, a DMFT solution involves $(i)$ replacing the 
lattice model by a self-consistently embedded multi-orbital, asymmetric 
Anderson impurity model, and $(ii)$ a selfconsistency condition which 
requires the impurity propagator to be equal to the local ($k$-averaged) 
Green function of the lattice, given by

\be
G_{\alpha}(\omega)=\frac{1}{V_{B}}\int d^{3}k \left[\frac{1}{(\omega+\mu)1
-H_{LSDA}^{0}({\bf k})-\Sigma(\omega)}\right]_{\alpha} \;.
\ee
Using the locality of $\Sigma_{\alpha\beta}$ in $d=\infty$, we have 
$G_{\alpha}(\omega)=G_{\alpha}^{0}(\omega-\Sigma_{\alpha}(\omega))$ from the 
Hilbert transform of the LDA DOS.  Also, importantly, the inter-orbital 
couplings scatter electrons between the $a_{1g},e_{g}^{\pi}$ bands, so that 
only the total number, $n_{t_{2g}}=\sum_{\alpha}n_{t_{2g},\alpha}$ 
is conserved in a way consistent with Luttinger's theorem.

To solve the multi-orbital, asymmetric Anderson impurity problem, we use the
iterated perturbation theory (IPT), suitably generalized to the case of 
$t_{2g}$ orbitals for arbitrary filling~\cite{[15]}. The local propagators 
are given by

\be
G_{\alpha}(\omega)=\frac{1}{N}\sum_{{\bf k}}\frac{1}{\omega
-\Sigma_{\alpha}(\omega)-\epsilon_{{\bf k}\alpha}} \;.
\ee
Local self-energies $\Sigma_{\alpha}(\omega)$ are computed within an extended
IPT scheme that explicitly satisfies the generalized Friedel sum rule 
(Luttinger's theorem) to a very good accuracy.  Mathematically,

\be
\Sigma_{\alpha}(\omega)=\frac{\sum_{\gamma}A_{\alpha\gamma}
\Sigma_{\alpha\gamma}^{(2)}(\omega)}{1-\sum_{\gamma}B_{\alpha\gamma}
\Sigma_{\alpha\gamma}^{(2)}(\omega)}
\ee
where, for example,

\be
\Sigma_{\alpha\gamma}^{(2)}(i\omega)=N_{\alpha\gamma}\frac{U_{\alpha\gamma}^2}
{\beta^2}\sum_{n m}G_{\alpha}^{0}(i\omega_{n})G_{\gamma}^{0}(i\omega_{m})
G_{\gamma}^{0}(i\omega_{n}+i\omega_{m}-i\omega)
\ee
with $N_{\alpha\gamma}=2$ for $\alpha,\gamma=e_{g1}^{\pi}, e_{g2}^{\pi}$ and 
$4$ for $\alpha,\gamma=a_{1g},e_{g1,2}^{\pi}$.  The bath propagator is 
$G_{\alpha}^{0}(\omega)=[\omega+\mu_{\alpha}-\Delta_{\alpha}(\omega)]^{-1}$.
In the above, $A_{\alpha\gamma}=\frac{n_{\alpha}(1-2n_{\alpha})+
D_{\alpha\gamma}[n]}{n_{\alpha}^{0}(1-n_{\alpha}^{0})}$ and 
$B_{\alpha\gamma}=\frac{(1-2n_{\alpha})U_{\alpha\gamma}+\epsilon_{\alpha}-
\mu_{\alpha}}{2U_{\alpha\gamma}^{2}n_{\alpha}^{0}(1-n_{\alpha}^{0})}$.  
Also, $n_{\alpha}$ and $n_{\alpha}^{0}$ are particle numbers 
defined from $G_{\alpha}$ and $G_{\alpha}^{0}$, and the inter-orbital 
correlation function is $D_{\alpha\gamma}[n]=\langle n_{\alpha}
n_{\gamma}\rangle=\langle n_{\alpha}\rangle\langle 
n_{\gamma}\rangle-\frac{1}{U_{\alpha\gamma}\pi}
\int_{-\infty}^{+\infty}f(\omega)Im[\Sigma_{\alpha}(\omega)
G_{\alpha}(\omega)]d\omega$.  This last identity follows directly from 
the equation of motion for $G_{\alpha}(\omega)$. 

These coupled equations are solved selfconsistently to obtain the spectral
function.  We choose $\Delta_{trg} \simeq 0.32~eV$, completely consistent with 
the LDA.  To study the MIT, we notice that external pressure decreases the 
quantity $\Delta_{trg}=E_{a_{1g}}-E_{e_{g}^{\pi}}$ and leads to an increase in 
$n_{a_{1g}}$; in particular, it leads to a reduction in $\Delta_{trg}$ across 
$0.28$, at which point, a sudden increase in $n_{a_{1g}}$ has been reported 
from cluster calculations~\cite{[7]}.  To study this effect, we monitor the 
spectral function for different values of $n_{a_{1g}}$, with fixed total 
number of electrons.

\begin{figure}[hb]
\epsfxsize=3.4in
\epsffile{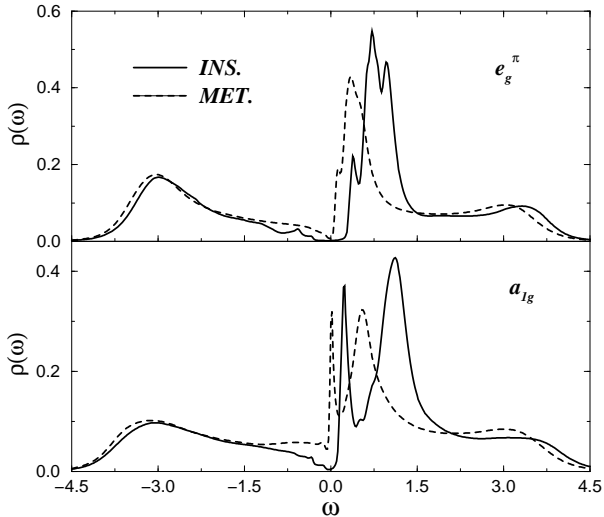}
\caption{$t_{2g}$ partial density of states using LDA+DMFT for $U=6.0~eV$ 
in the PI (solid) and PM (dashed) phases of $V_{2}O_{3}$.} 
\label{fig2}
\end{figure}

In Fig.~\ref{fig2}, we show the partial DOS for the $e_{g}^{\pi}, a_{1g}$
orbitals, for the metallic (dashed) and ``Mott'' insulating (solid) phases, 
corresponding to total $a_{1g}$ occupation $n_{a_{1g}}=0.41,0.36$, 
respectively.  As expected from the trigonal splitting assignment, the 
lower-lying $e_{g}^{\pi}$ orbitals are more localized in the solid, with a 
gap, $\Delta_{e_{g}^{\pi}}=0.45~eV$.  Correspondingly, the $a_{1g}$ band is 
more itinerant, completely consistent with the fact that $t_{a_{1g},a_{1g}}$ 
{\it is} by far the largest hopping integral in the real system.  From 
Fig.~\ref{fig2}, we estimate $\Delta_{Ins}=0.35~eV$. With $n_{a_{1g}}=0.41$, 
the $e_{g}^{\pi}$ spectral function still exhibits insulator-like features, 
while the $a_{1g}$ spectrum develops a very narrow, quasicoherent peak with 
a $FWHM=0.07~eV$ at $E_{F}$. We interpret these results as a microscopic 
derivation of the phenomenological ``two-fluid'' models proposed earlier 
to understand metal-insulator transitions~\cite{[3]}.  In our new picture, 
increasing pressure decreases $\Delta_{trg}$, resulting in increased population
of the $a_{1g}$ orbital, and leading (via $t_{a_{1g},a_{1g}}$) to an abrupt 
MIT accompanied by large dyna-
\begin{figure}[htb]
\epsfxsize=3.4in
\epsffile{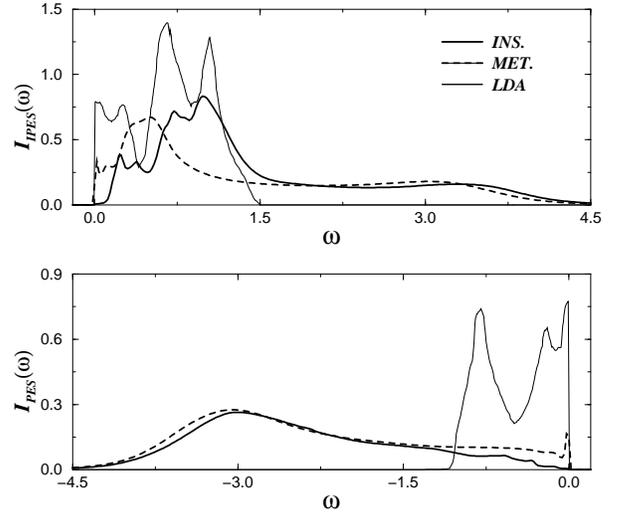}
\caption{The integrated inverse- and direct photoemission lineshapes for 
the PI (solid) and PM (dashed) phases within LDA+DMFT for $U=6.0~eV$.  The 
LDA result is the thin solid line.} 
\label{fig3}
\end{figure}
\hspace{-0.35cm}mical SWT.

Using the corresponding self-energy, we carry out a quantitative estimate 
for the effective mass enhancement and the $A$-coefficient of the low-$T$
quadratic term in the $dc$ resistivity. Indeed, from 
$Re[\Sigma_{a_{1g}}(\omega)]$, we estimate $m^{*}/m=[1-
\frac{\partial\Sigma_{a_{1g}}^{'}(\omega)}{\partial\omega}]^{-1}=4.16$, 
close to the value of $4.4$ extracted from low-$T$ specific heat measurements. 
Also, $A=\rho_{dc}(T)/T^{2}=(m^{*}/ne^{2})|{\partial^{2}
\Sigma_{a_{1g}}^{''}(\omega)}/{\partial\omega^{2}}|$ is large, because
$|{\partial^{2}\Sigma_{a_{1g}}^{''}(\omega)}/{\partial\omega^{2}}|=38\gg 1$.  
This will result in a large quadratic term in the low-$T$ $dc$ 
resistivity, as observed.  

In Fig.~\ref{fig3}, we show the integrated PES and IPES spectra in the 
metallic and insulating phases, along with the corresponding LDA spectra. As 
one would expect, and consistent with observations, the PES spectrum shows 
a very narrow quasicoherent peak, with most of the spectral weight in the 
incoherent ``lower Hubbard band'', centered at $\omega=-3.0~eV$. The PES 
intensity at $E_{F}$ is strongly reduced from the LDA prediction.  
Quantitative comparison with the PES spectra measured at higher $T$
 needs an extension of our approach to finite $T$; however, we expect that the
strong $T$-dependence of the quasicoherent feature within DMFT will smear out
the low-energy peak for $T>T_{coh}\simeq 300K$, bringing the spectra in closer 
agreement with published PES results.  Signatures of strong correlations are
also visible in the IPES spectra as ``upper Hubbard band'' features, and the 
difference between the LDA and LDA+DMFT spectra is striking. 

\begin{figure}[htb]
\epsfxsize=3.4in
\epsffile{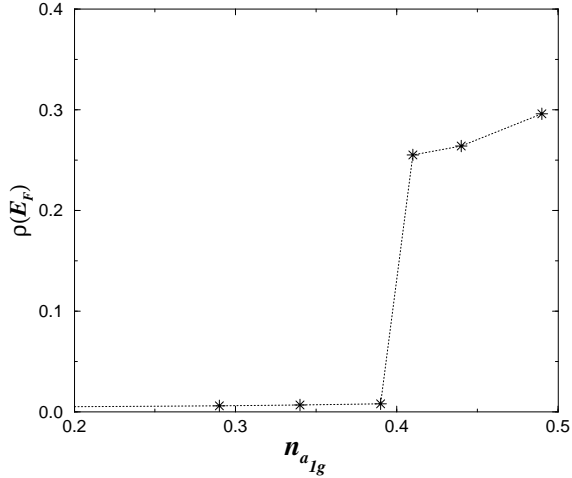}
\caption{ The total DOS at $E_{F}$ as a function of the $a_{1g}$ orbital 
occupation.} 
\label{fig4}
\end{figure}

Especially interesting is the change in the spectral function at $E_{F}$ as 
a function of $n_{a_{1g}}$. In Fig.~\ref{fig4}, we show this variation 
for our chosen parameter set. Astonishingly, $\rho(E_{F})$ exhibits 
a sharp jump from $0$ to $0.261$
\hspace{-0.35cm}around $n_{a_{1g}}=0.39$, clearly showing that the MIT is 
indeed first order, in agreement with observations. Our results imply that 
the MIT is accompanied by an abrupt jump in the itinerant carrier density, 
rather than increase in carrier mobility, and is consistent with the 
conductivity jump at the PI-PM transition. From the computed spectral 
functions, $(n_{a_{1g}},n_{e_{g1}^{\pi}},n_{e_{g2}^{\pi}})= (0.36,0.82,0.82)$ 
in the PI, and $(0.41,0.79,0.79)$ in the PM. Our calculated orbital 
occupations are slightly different from those estimated by polarized XAS 
measurements~\cite{[5]}, but show the correct trend across the MIT. Further, 
we find $\langle S_{iz}^{2}\rangle=\frac{1}{4}\sum_{\alpha}\langle 
[n_{i\alpha\uparrow}-n_{i\alpha\downarrow}]^{2}\rangle=0.92$ for the chosen 
parameter set, consistent with the observation of $S=1$ at each $V$ site. 
Finally, we monitor 
$D_{a_{1g},e_{gi}^{\pi}}$ and $D_{e_{g1}^{\pi},e_{g2}^{\pi}}$ and 
find that both {\it decrease} across the MIT: in fact, 
$D_{a_{1g},e_{gi}^{\pi}}=-0.043$ for $n_{a_{1g}}=0.36$ (PI), and equals 
$-0.007$ for $n_{a_{1g}}=0.41$ (PM), while 
$D_{e_{g1}^{\pi},e_{g2}^{\pi}}=-0.065$ (PI) and $-0.036$ (PM) showing clearly 
that the MIT is accompanied by a strong {\it reduction} of local (anti-ferro) 
orbital correlations.

As mentioned earlier, the MIT in $V_{2}O_{3}$ has recently been studied
~\cite{[13]} by LDA+DMFT(QMC), so it is instructive to compare our approach 
with theirs.  Though QMC is numerically exact, it cannot access the low-$T$ 
regime ($T<400K$).  Further, analytic continuation of Matsubara (imaginary) 
frequencies onto the real frequencies is a highly ill-posed numerical problem.
While the IPT is not exact, it is free from these drawbacks, and is known to 
yield reliable results for the problem at hand~\cite{[15]}, enabling one to 
study thermodynamic and transport properties at relatively modest computational
cost.  In contrast to the earlier pioneering study, we have explicitly shown
the existence of the first-order PI-PM transition and its intimate link to the 
accompanying orbital switching in $V_{2}O_{3}$.  We have not attempted to make 
a detailed comparison with finite-$T>170K$ spectral functions in the PI and PM 
regions, a program which has been carried out in Ref.~\cite{[13]} (but only 
for $T>400~K$, where the first-order MIT is replaced by a smooth crossover), 
and we plan to report these aspects, along with finite-$T$ transport 
properties and lattice effects, in a future work.  

In conclusion, we have presented a different scenario, based on combination 
of LDA bandstructure with dynamical, local spin and orbital correlations, to 
understand salient features of the first-order MIT in the P-phase of 
$V_{2}O_{3}$. The first-order MIT is shown to be accompanied by changes in 
orbital occupation in a way qualitatively consistent with observations.  
Further, very good agreement with low-$T$ thermodynamic and $dc$ resistivity 
is obtained within the same approach.  This represents a new picture for the 
correlation-driven MIT in $V_{2}O_{3}$; one which has applications to other 
systems where coupled spin-orbital correlations result in changes in orbital 
occupations across the MIT. Such an approach should be applicable, with 
extensions to include broken symmetries in the spin/orbital sectors, to 
systems like $Ca_{2-x}Sr_{x}RuO_{4}$ as well. 

We are indebted to L. H. Tjeng for many enlightening discussions.  Work 
carried out with the support of the Sonderforschungsbereich 608 of the 
Deutsche Forschungsgemeinschaft.

\end{document}